\input stromlo

\title Near-Infrared Synthetic Spectra of Elliptical Galaxies

\shorttitle Near-IR Synthetic Spectra

\author Mark L. Houdashelt and Roger A. Bell

\shortauthor Houdashelt \& Bell

\affil Department of Astronomy, University of Maryland

\abstract We present near-infrared (0.7--3 $\mu$m) synthetic spectra from
our initial population synthesis models of elliptical galaxies.  We measure
broad-band colors 
from the spectra and compare the results to the models of Worthey (1994) and
to observations of E/S0 galaxies. 

\section Introduction

In theory, knowledge of the present-day stellar populations of a galaxy can be used to
derive the star formation history of that galaxy.  In practice, however,
decomposing the integrated light of a galaxy to determine the characteristics
of its various stellar
populations is a daunting task.  Even in a simple stellar population (SSP),
{\it i.e.} a coeval group of stars having a common chemical composition, the
effects of age and metallicity variations on the integrated light can be
quite degenerate (see Worthey 1994; hereafter W94).

One analytical method which is a potentially powerful tool in integrated light
studies is population synthesis.  With this application in mind,
we have used synthetic stellar spectra to construct integrated
spectra of elliptical galaxies.  We present here our initial near-infrared
(0.7--3 $\mu$m) synthetic spectra of E galaxies, examine how these
spectra are affected by changes in age
and metallicity and compare our results to the models of W94.  This
research builds upon the previous work of
Tripicco \& Bell (1995), Houdashelt \& Bell (1996), Bell \& Briley (1991)
and Houdashelt (1995).  A companion paper in this volume presents a
complementary discussion of our optical synthetic spectra (3000--7000 \AA).

\section Some Details of the Models

In this initial set of models, elliptical galaxies were represented by SSPs.
In this way, the effective temperature, surface gravity
distribution of the stars in a galaxy could be described by a single isochrone.
We used Dorman (1996, private communication) isochrones through the tip of the
first-ascent red giant branch (RGB) and Sweigart (1996, private communication)
horizontal-branch and asymptotic-giant-branch tracks thereafter.
A Salpeter (1955) IMF was assumed in each case.  The nine synthetic spectra
discussed here represent 6, 10 and 16 Gyr SSPs
having [Fe/H]=+0.39, 0.00 and --0.47.

Updated MARCS stellar atmosphere models (Gustafsson {\it et al.} 1975)
were calculated for a representative sample of stars along each isochrone,
and synthetic spectra were then computed using the SSG spectral synthesis code
(Bell \& Gustafsson 1978, 1989; Gustafsson \& Bell 1979).
In computing the synthetic spectra, the
microturbulent velocity, {\it v}$_t$, was varied linearly
with log {\it g}, such that (1) {\it v}$_t$=1.0 km sec$^{-1}$ for
log~{\it g}~$\ge$~4.5, (2) {\it v}$_t$=2.5 km sec$^{-1}$ for
log~{\it g}~$\le$~0.5, and (3) {\it v}$_t$=1.7 km sec$^{-1}$ for
log~{\it g}=1.7.  Solar abundance ratios were used in all but two instances.
For consistency with the isochrones, the [Fe/H]=--0.47 models also had
[O/Fe]=+0.23.  In addition, following observational determinations,
all stars which were more evolved than the RGB bump were assumed to have mixed
CNO-processed material into their stellar atmospheres, such
that [C/H]=--0.19, [N/H]=+0.28, [O/H]=--0.04, and $^{12}$C/$^{13}$C=14
(Smith \& Lambert 1990).

\section Results

Figure 1 illustrates the near-infrared synthetic spectrum of our 16 Gyr,
solar-metallicity elliptical galaxy.  The $JHK$ filter profiles of Bessell \&
Brett (1988) and the filter profiles of the CIT/CTIO CO index (Cohen
{\it et al.} 1978) are also plotted.
The well-known $\Delta${\it v}=3 and $\Delta${\it v}=2
CO bands with bandheads near 1.5 and 2.3 $\mu$m, respectively, are easily
discerned in the synthetic spectrum and are obviously the dominant features 
in the $H$ and $K$ bands.  Most of the strong absorption bands shortward
of 1.5 $\mu$m are due to TiO; the $\phi$ band of TiO is especially important
between $\sim$0.9 and 1.5 $\mu$m and has a significant impact on the $J$-band
flux.  The Ca\II\ triplet near 8500 \AA\ is evident in Figure 1 as well.
Since we are not aware of any medium- or high-resolution spectra
of E/S0 galaxies between 1 and 3 $\mu$m, the spectrum shown in Figure 1
can serve as a guide to observers until such data exists.

\figureps[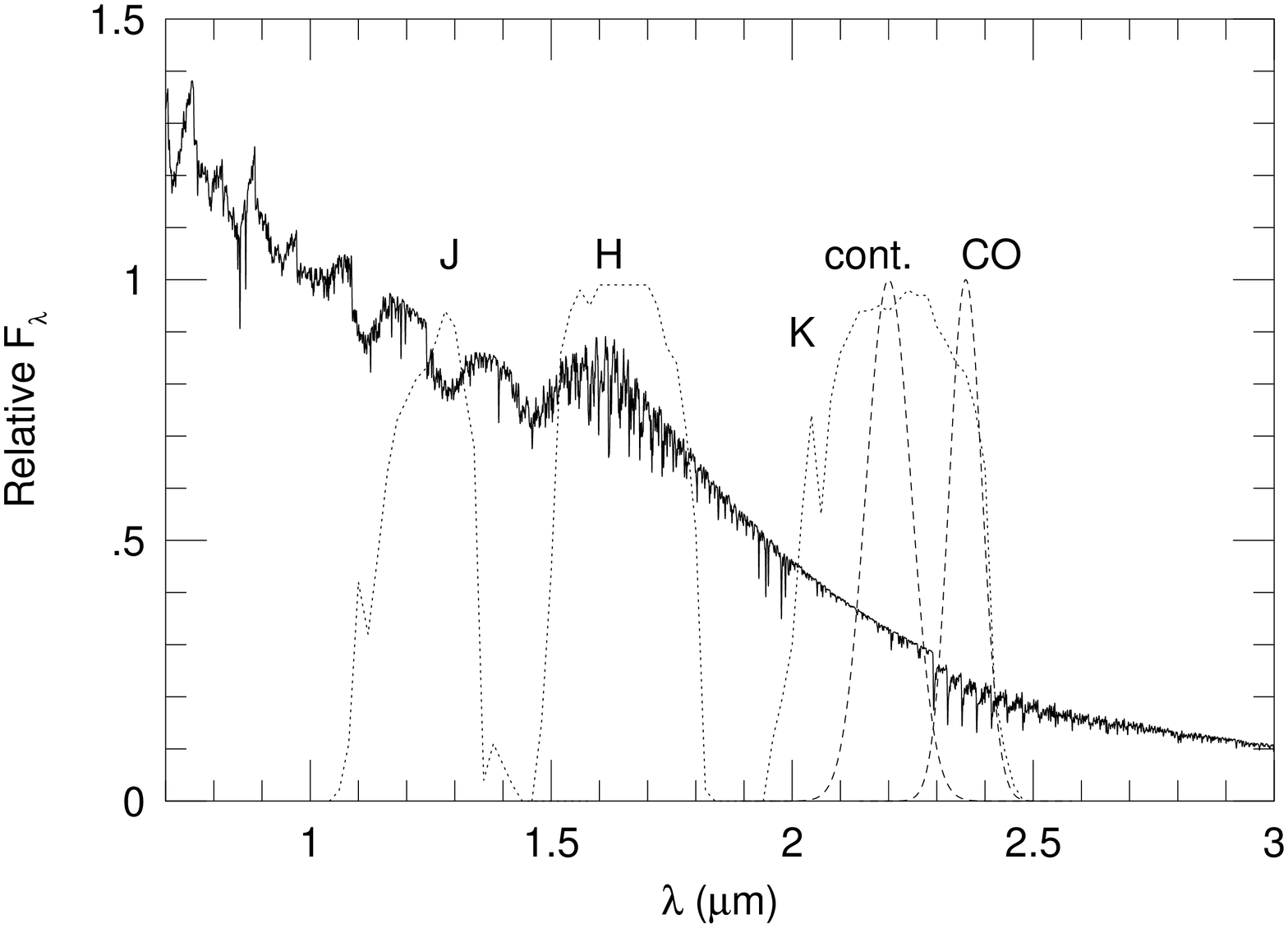,.8\hsize] 1. Our near-infrared synthetic spectrum for a
16 Gyr, [Fe/H]=0.0 stellar population (solid line).  The dotted lines are the
$J$, $H$ and $K$ filter profiles on the Bessell \& Brett (1988) Johnson-Glass
system.  The dashed lines are the filter profiles used to measure the CIT/CTIO
CO index (Cohen {\it et al.} 1978).

To evaluate our models, we measured integrated colors from the synthetic
spectra.  Figure 2 compares our near-infrared colors to observations of
early-type galaxies and to the model predictions of W94; all of the
photometry has been put on the Johnson-Glass system of Bessell \& Brett (1988).

\figureps[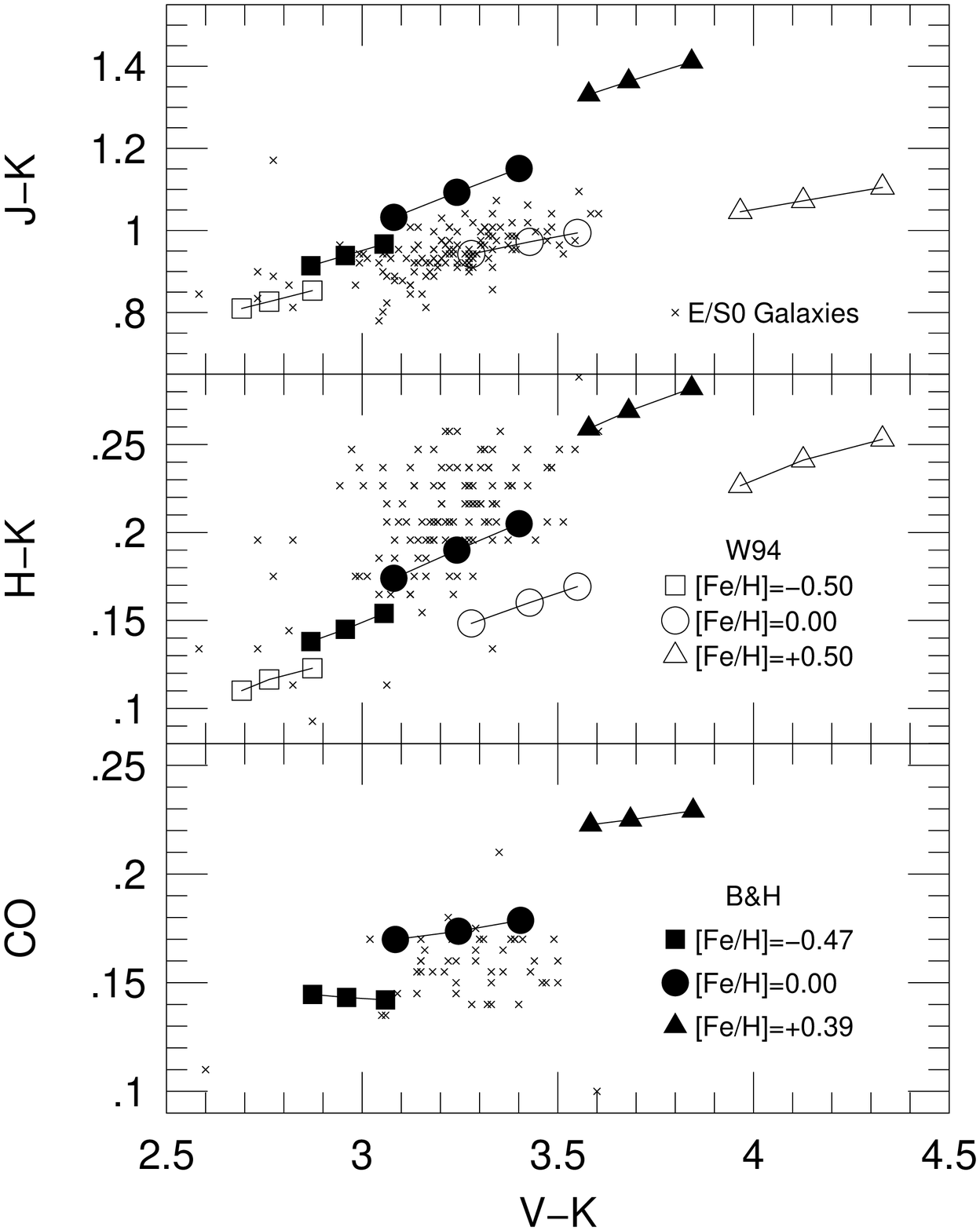,.6\hsize] 2. The near-infrared colors measured from
our spectral synthesis models are compared to the model predictions of W94
and the observed colors of early-type galaxies (Frogel {\it et al.} 1978,
Persson {\it et al.} 1979).
From blue to red, our models represent 6, 10 and 16 Gyr populations; the
W94 models shown have ages of 8, 12 and 17 Gyr.

Taken alone, the $V-K$ colors indicate that the reddest ellipticals are
most likely old and have approximately solar metallicity.  Whether the
bluer ellipticals are younger or more metal-poor cannot be determined from
the integrated colors, since the age and metallicity vectors are nearly
parallel in the color-color plots, as previously known (W94).
Since the $V-K$ colors are consistent
with the age and metallicity inferred from the $B-V$ colors of our models
(see Bell \& Houdashelt elsewhere in this volume), we are confident that
our synthetic spectra are accurately representing the $K$-band fluxes of
elliptical galaxies.

Silva (1996) has shown that the population synthesis models of W94 and
Bruzual \& Charlot (1993) cannot reproduce the $J-H$, $H-K$ colors of
early-type galaxies.  Figure 2 indicates that our present models suffer
from similar difficulties but with some differences.  While our models
reasonably match the E/S0 galaxies in the $H-K$, $V-K$ diagram and the W94
models do not, the opposite is true for the $J-K$, $V-K$ colors.
However, comparisons of our synthetic stellar spectra and spectra
of field M giants from Terndrup {\it et al.} (1991) lead us to believe that
we have overestimated the oscillator strengths of the spectral lines in the
$\phi$ system of TiO, which is very
prominent in the synthetic spectrum shown in Figure 1.
The $J$-band magnitudes are particularly sensitive to this system of
TiO, so TiO bands which are too strong would make our $J-K$ colors too red.
We intend to explore this hypothesis more fully in the near future, but
for now, a significant discrepancy remains between our model $J-K$ colors and
those of early-type galaxies.  On the other hand,
our model $V-K$, $H-K$ trends are a much better match to the galaxy data,
even though the E/S0 galaxies tend to be slightly redder in $H-K$ than our
solar-metallicity models.

We have had great success in synthesizing the 2.3 $\mu$m CO band in giants
(Briley \& Bell 1991, Houdashelt \& Bell 1996), and
our model CO indices span the observational
measurements of early-type galaxies.  While our models infer that giant
ellipticals have slightly subsolar metallicities, we must
explore the sensitivity of the CO bands to our treatment of CNO mixing
in the giants before drawing any specific conclusions from the CO indices.

This work was supported by NASA Grant NAG53028 and NSF Grant AST93--14931.

\references

Bell, R.A. \& Briley, M.M. 1991, AJ, 102, 763.

Bell, R.A. \& Gustafsson, B. 1978, A\&AS, 34, 229.

Bell, R.A. \& Gustafsson, B. 1989, MNRAS, 236, 653.

Bessell, M.S. \& Brett, J.M. 1988, PASP, 100, 1134.

Bruzual, A.G. \& Charlot, S. 1993, ApJ, 405, 538.

Cohen, J.G., Frogel, J.A. \& Persson, S.E. 1978, 222, 165.

Frogel, J.A., Persson, S.E., Aaronson, M. \& Matthews, K. 1978, ApJ, 220, 75.

Gustafsson, B. \& Bell, R.A. 1979, A\&A, 74, 313.

Gustafsson, B., Bell, R.A., Eriksson, K. \& Nordlund, A. 1975, A\&Ap, 42, 407.

Houdashelt, M.L. 1995, Ph.D. Dissertation, The Ohio State University.

Houdashelt, M.L. \& Bell, R.A. 1996, BAAS, 28, 915.

Persson, S.E., Frogel, J.A. \& Aaronson, M. 1979, ApJS, 39, 61.

Salpeter, E.E. 1955, ApJ, 121, 161.

Silva, D.R. 1996, in {\it Spiral Galaxies in the Near-IR}, ed. D. Minnitti \&
H.-W. Rix (Springer-Verlag, Berlin), p. 3.

Smith, V.V. \& Lambert, D.L. 1990, ApJS, 72, 387.

Terndrup, D.M., Frogel, J.A. \& Whitford, A.E. 1991, ApJ, 378, 742.

Tripicco, M.J. \& Bell, R.A. 1995, AJ, 110, 3035.

Wing, R.F. \& Ford, K.F. 1969, PASP, 81, 527.

Worthey, G. 1994, ApJS, 95, 107. (W94)

\bye